# Formation of molecular rectifier with gold nanoclusters

Sudipta Pal [1], Milan K. Sanyal [1], Neena S. John [2], Giridhar U. Kulkarni [2]

[1]*Surface Physics Division, Saha Institute of Nuclear Physics, 1/AF, Bidhannagar, Kolkata-700 064, India.*

[2] *Chemistry and Physics of Materials Unit, Jawaharlal Nehru Center for Advanced Scientific Research, Jakkur, Bangalore-560064, India.*

**Gold nanoclusters encapsulated with organic molecules are of great interest for its possible applications in various fields of nanotechnology like molecular electronics, catalysis and medical science. Here we demonstrate that monolayer and bilayer films of thiol-capped gold nanoclusters can exhibit diode-like properties provided controlled spatial asymmetry is created between two tunnel junctions used to connect a gold nanocluster *molecule*. Current-voltage characteristics of this molecular rectifier were obtained from conducting probe atomic force microscopy measurements and also from conventional two probe resistance measurements. Systematic x-ray reflectivity and atomic force microscopy measurements were carried out to characterize the spatial asymmetry introduced by a monolayer of fatty acid salt gadolinium stearate used to deposit thiol-capped gold nanocluster molecules on hydrophilic $SiO_2$-Si(001) substrate by Langmuir Blodgett technique. The prominent rectification property observed in these nano-structured films could be explained from the measured spatial asymmetry. An obvious application of this work will be to form an array of memory *dots* with gold nanoclusters by depositing these on a monolayer of organic molecules having desired "tail length".**

It is quite obvious that the molecular rectifier (MR) will be an important ingredient in nanotechnology and the simplest application of MR will be in memory devices with enormous storage capacity [1]. The present status of this explosively growing field is far from the MR suitable for technological applications that need sharp voltage thresholds, large current rectification ratios, small time constants and above all easy-to-form stable molecular structures. Experimental and theoretical studies of current-voltage characteristics of monomolecular films are being carried out extensively to achieve these desired properties of MR and these studies also improve our basic understanding in the electrical transport phenomena in low dimension. The original concept [2] of MR was developed with the assumption that highest occupied molecular orbital (HOMO) and lowest unoccupied molecular orbital (LUMO) can be confined in two parts of a molecule separated by an insulator bridge that prevent orbital overlap or "spilling off" forming a donor-insulator-acceptor (D-$\sigma$-A) structure. Monomolecular film of a zwitterionic molecule grown by Langmuir-Blodgett (LB) technique exhibited properties [3] of MR and was thought to be an experimental verification of the molecular D-$\sigma$-A concept. It is now realized [4] that having an insulator ($\sigma$) bridge within a molecule may be questionable in most cases and one can form MR easily by connecting a molecule asymmetrically to two tunnel barriers from each side in a line-of-sight. In this situation one can in principle design a MR by altering spatial asymmetry of the connecting tunnel barriers [1,5]. Here we demonstrate that one can form MR using dodecanethiol capped gold nanocluster of 3.3 nm diameter by forming asymmetric spatial configuration — one side having thiol chain and the otherside silicon oxide and a fatty acid salt, gadolinium stearate. Although electrical properties of metal nanoclusters connected by organic tails are being studied rigorously [6-9], to the best of our knowledge the MR formation through spatial asymmetry in these well-studied Au nanocluster systems has not been reported earlier.

A schematic diagram of a typical film is shown in Fig. 1(a) with the electrical connection used for direct current-voltage (*I-V*) measurements in a conducting probe atomic force microscope (C-AFM). The films were deposited in an LB trough on a native oxide covered doped Si (001) substrate through a sequential process involving two water-to-air strokes. The substrate cleaning and other details have been described earlier [10]. In the first stroke the substrate was taken through a compressed Langmuir monolayer of stearic acid having gadolinium ions in the water subphase forming gadolinium stearate (GdSt) monolayer on hydrophilic $SiO_2$-Si substrate. Then this hydrophobic GdSt-$SiO_2$-Si substrate was again taken from water-to-air through a Langmuir monolayer of dodecanethiol encapsulated Au nanoclusters forming the second monolayer on GdSt. We have already shown [10] that the presence of GdSt monolayer improves the film quality and third monolayer can grow if the surface pressure of the Langmuir monolayer of Au nanoclusters is kept higher. We present here results of two types of films marked as A and B, deposited with surface pressures of 1.5 and 20.0 mN/m respectively, for investigating electrical properties of film having monolayer and bilayer of Au nanoclusters on GdSt-$SiO_2$-Si substrate. We used x-ray reflectivity and AFM measurements of GdSt-$SiO_2$-Si substrate and Au nanoclusters covered films to determine thicknesses accurately as spatial configuration is the key feature here. The I-V characteristics of the bare $SiO_2$-Si, GdSt-$SiO_2$-Si substrate and films A and B were measured by a C-AFM setup (Nanoscope IV, Digital Instruments) with a gold coated $Si_3N_4$ tip. We also cross-checked I-V characteristics of GdSt-$SiO_2$-Si system by approaching the C-AFM tip in the "holes" (Tip position 2 in Fig. 1a) while measuring films A and B to ensure that characteristic of GdSt-$SiO_2$-Si structure had remained almost unaltered if it is not covered with Au nanoclusters in the last stroke of LB deposition. We have also measured *I-V* characteristic of the films by two-probe technique with an electrometer (Keithley 6517A) to confirm diode-like properties observed in C-AFM measurements.

In Fig. 1(b) we have shown x-ray reflectivity profiles of film A and B and GdSt-$SiO_2$-Si substrates. The electron density profiles (shown in the insets) are extracted by fitting [10] and the fitted reflectivity profiles are also shown with measured data. Electron density profiles of the films and measured AFM morphology of the films (also shown in the insets) were used to determine [10] the distances accurately as indicated in the simplified model shown in Fig. 1(a). With native $SiO_2$ thickness of 20Å asymmetric spatial tunnel barrier is apparent even in film A having primarily one Au cluster layer on GdSt-$SiO_2$-Si substrate. The C-AFM tip is about 11Å away from the Au cluster and the other conducting connection of doped Si(001) is around 32Å (=20Å for $SiO_2$ + 12Å for thiols and GdSt layer) away. This spatial asymmetry gives rise to diode-like *I-V* characteristics in C-AFM measurements provided the tip is placed on the top of an Au-cluster. It is to be noted here that the separation (12Å) of Au cluster and $SiO_2$-Si interface is much less than expected thickness of thiol plus stearic acid tails (14Å+20Å=34Å) indicating rearrangements of organic tails below the Au cluster.

We have shown a C-AFM tip position (point 1 in Fig. 2a and 1a) on film A and corresponding *I-V* characteristics of this Au cluster in Fig 2(b) as conductance (d*I*/d*V*) plot as a function of voltage applied to the tip. We observe sharp rise in conductance beyond a forward voltage ($V_F$) of + 240mV but reverse voltage $V_R$ comes out to be around -450 mV for obtaining non-zero conductance. At ± 500mV ratio of forward to reverse current defined as current rectification ratio (*R*) comes out to be 13.5. It is interesting to note that we get symmetric conductance vs. voltage curve and *R* as unity if the tip is not placed on the Au cluster (point 2 in Fig. 2a and 1a). The tip is now on GdSt layer as also evident in height profile of Fig. 2a and obtained conductance vs. voltage is similar to that obtained by measuring bare GdSt-$SiO_2$-Si substrate shown in Fig. 2b. This observation also indicates that tail distortion takes place only below an Au cluster. In Fig. 2b we have also shown measured conductance spectra of bare $SiO_2$ covered doped Si(001) substrate used in this experiment. The symmetric profile observed here is typical tunneling data through $SiO_2$ barrier that depend strongly on tip separation [5]. The spatial asymmetry becomes more prominent for film B provided the tip is placed on a bilayer of Au nanocluster (point 3 in Fig. 2a and 1a). In this condition we could not get conductance in the negative voltage side and conductance in the positive voltage side remains similar but slightly better (Fig. 2b) and *R* comes out to be 364.8. In this film also we get back conductance profile (refer Fig. 2b) of GdSt-$SiO_2$-Si if the tip is placed on GdSt (point 4 in Fig. 2a) that is around 90Å below the top Au nanocluster as shown in height profile. Moreover we get back conductance profile of Au nanocluster of film A by positioning the tip (point 1 in Fig. 1a) on an Au cluster situated in first monolayer of film B. This result confirms that in-plane conduction through Au nanoclusters is not dominating here. The results in Fig. 2b clearly demonstrate that dominating current carrying paths here is tunnelling through C-AFM tip -to -Au nanocluster-to- doped Si structure and one can produce molecular rectifiers even with thiol capped Au clusters having diameter of 3.3 nm.

This rectifying characteristics of films A and B is not observed when the connection is taken from the film itself avoiding tunneling barrier of $SiO_2$ as we have reported earlier [10]. We present here standard two-probe conductance data to illustrate the role of asymmetry further. We took two probe *I-V* data with the electrometer by making one connection with a conducting pad from the top of the film — the other connection was taken either from another portion of film (curve 4 in Fig. 2c) or from the back of the Si substrate (curve 2 in Fig. 2c). The curve 1 in Fig. 2(c) is same as the data of point 1 shown in Fig. 2(b) and can be used as reference. The rectification is evident even in two-probe data when connection is taken from back of the substrate (curve 2 of Fig. 2c) but contrast is less as the conducting pad averages the conductance of Au nanocluster with that of GdSt due to patchy nature of Au cluster monolayer (refer Fig. 2a). It is interesting to note that the rectification feature does not appear in both C-AFM and two-probe data shown as curve 3 and 4 respectively in Fig. 2(c) when the other connection is taken from the film avoiding $SiO_2$ interface.

In Fig. 3 we have explained MR property of Au nanocluster observed here on the basis of spatial asymmetry in two tunnel barriers — one formed by organic tails of thiol molecules lying between C-AFM tip and Au nanocluster (top barrier of thickness $L_t$=11Å) and the other formed by thiol-GdSt-$SiO_2$ barrier in between Au nanocluster and Si (bottom barrier of thickness $L_b$=32Å). The applied voltage, V, drops asymmetrically on top and bottom barriers giving rise to MR property as expected [1,4,5]. Let the voltage drop at bottom and top barrier be $V_b$ and $V_t$ respectively with their ratio $\eta = V_b/V_t$. Here we should mention that in the schematic diagram we do not show the band gap of the doped Si(001) explicitly because in GdSt-$SiO_2$-Si system the effect of band bending and the presence of surface states in the Si gap region make it difficult to locate the valence and conduction band edge accurately with respect to the Fermi level. So to the first approximation we treat the heavily doped Si as a metallic electrode. At equilibrium the Fermi levels of C-AFM tip and Si align with the Fermi level of Au nanoparticle as shown in Fig 3a. The equilibrium Fermi level position obviously depends on the relative work functions of the two electrodes, which are not same in our case. The difference in work functions induces a contact potential difference and hence an electric field through the system at zero applied bias. For simplicity we ignore the field within the nanocluster and assume that the applied voltage drops only on the two barriers i.e. $V=V_b + V_t$. In our diagram, the bottom electrode (Si) is always grounded and all the energy levels shift with respect to Si Fermi level with applied bias. From density functional study it has been shown earlier [11] that a thiol passivated Au cluster consisting of 38 Au atoms has a gap of 0.9 eV just above the Fermi energy. We denote this gap as $\Delta=(E_{vac} - E_f) - A_1$ where $E_{vac}$, $E_f$ and $A_1$ are the vacuum level, equilibrium Fermi level position and electron affinity of Au nanoparticle at zero bias respectively. When the positive voltage is applied to the tip, the tip Fermi energy and the energy levels of Au nanocluster shift downward, but their relative shifting will depend on $\eta$. At a particular forward bias voltage ($V_F$) Si Fermi level lines up with the conduction levels of Au nanocluster (Fig. 3b) and a sharp increase in current is observed in *I-V* spectra due to resonant tunnelling [4,5]. Similarly, in reverse bias both the tip Fermi energy and the energy levels of Au cluster shift upwards and current will start to increase at a voltage ($V_R$) only when tip Fermi level lines up with the conduction levels of Au cluster (Fig. 3b). From the expressions [4] $eV_F = \Delta\,(1+\eta)\,/\,\eta$ and $eV_R = \Delta\,(1+\eta)$ , the asymmetric factor $\eta = V_R/V_F$ is found to be 1.9 for sample A. For sample B, current is almost zero in negative bias giving $\eta$>>1. From the above expressions $\Delta$ is found out to be 155 meV. This is consistent with earlier reports [12,13] which mentioned that HOMO-LUMO gap present in a thiol-capped Au nanocluster of 0.5 and 2 nm diameter is 1.8 and 0.3eV respectively. Here we should mention that the asymmetric parameter (denoted as $\eta'$ here) first introduced by Tian et al.[5] differs from the definition of $\eta$ used here [4]. From definition one can easily relate $\eta$ with $\eta'$ ($=V_t/V$) as $\eta = (1-\eta')/\,\eta'$ and $\eta'$ is found to be 0.34 for sample A. The asymmetric factor $\eta$ can be approximated [5,14] in terms of the parallel plate capacitances of the two junctions using a simple dielectric model with the widths and relative dielectric constants of top and bottom barriers $L_t$, $L_b$ $\varepsilon_t$ and $\varepsilon_b$ respectively. Assuming further that the area associated with the tunneling phenomena for the two barriers are same, one can write $\eta = (L_b/L_t)(\varepsilon_t/\varepsilon_b)$. Considering $L_b$ = 32Å, $L_t$ =11Å and $\varepsilon_t$ =2.2, which is the commonly used [7] value for alkane chains, $\varepsilon_b$ comes out to be 3.3. The dielectric constant of the bottom barrier is expected to be higher than the top one because of the presence of $SiO_2$ . It is to be noted here that by keeping only organic tails on both sides of Au nanoclusters one can get better $\eta$ and hence better *R*. Obvious extension of our finding is that one can create an array of MR by depositing thiol-capped Au nanocluster on an organic monolayer attached to a doped silicon or metal substrate. The *R* factor of the MR can then be designed by choosing the “tail length” of the organic monolayer ($L_b$).

The authors would like to acknowledge the help of Mr. Atikur Rahman in two probe measurements. Authors are also grateful to Prof. C. N. R. Rao for his support in this collaboration work.

Correspondence and requests for materials should be addressed to Milan K. Sanyal (e-mail: milan@lotus.saha.ernet.in).

Fig. 1: (a) Schematic diagram of the Au nanocluster film on GdSt-$SiO_2$-Si substrate along with the electrical connection for C-AFM measurements. Three tip positions are shown on the monolayer of Au nanoparticles (1), on GdSt-$SiO_2$-Si substrate (2) and on bilayer of Au nanoparticles (3). (b) X-ray reflectivity data (scatter graphs) along with the fitted profiles (solid lines) for sample A, B and GdSt-$SiO_2$-Si substrate. Corresponding electron density profiles (EDPs) and tapping mode AFM images are shown in the insets. The length scales obtained from x-ray analysis are marked in EDPs and also in (a). Top and bottom barrier widths ($L_t$ and $L_b$ respectively) are shown in EDP of film A.

Fig. 2: (a) AFM images of film A and B along with the corresponding height profiles. The tip positions for *I*-*V* measurements are marked in the images and also in the profiles. (b) d*I*/d*V* vs. *V* plots for sample A, B, GdSt-$SiO_2$-Si substrate and bare $SiO_2$-Si. Four curves are shown for sample A and B corresponding to four marked tip positions. (c) d*I*/d*V* vs. V plots obtained from C-AFM and two probe measurements by taking the second connections from backside of Si substrate (curve 1 and 2) and from film (curve 3 and 4) [refer text for details].

Fig. 3: (a) Schematic diagram showing the asymmetric tunnel barriers. Fermi levels are aligned at zero bias. (b) At a positive bias ($V_F$) to tip, Si Fermi level lines up with conduction level of Au nanocluster. (c) At a negative bias ($V_R$) to tip, tip Fermi level aligns with conduction level of Au cluster. The voltage drops across top and bottom barriers and electron affinity of Au cluster are marked as $V_t$, $V_b$ and $A_i$ respectively.

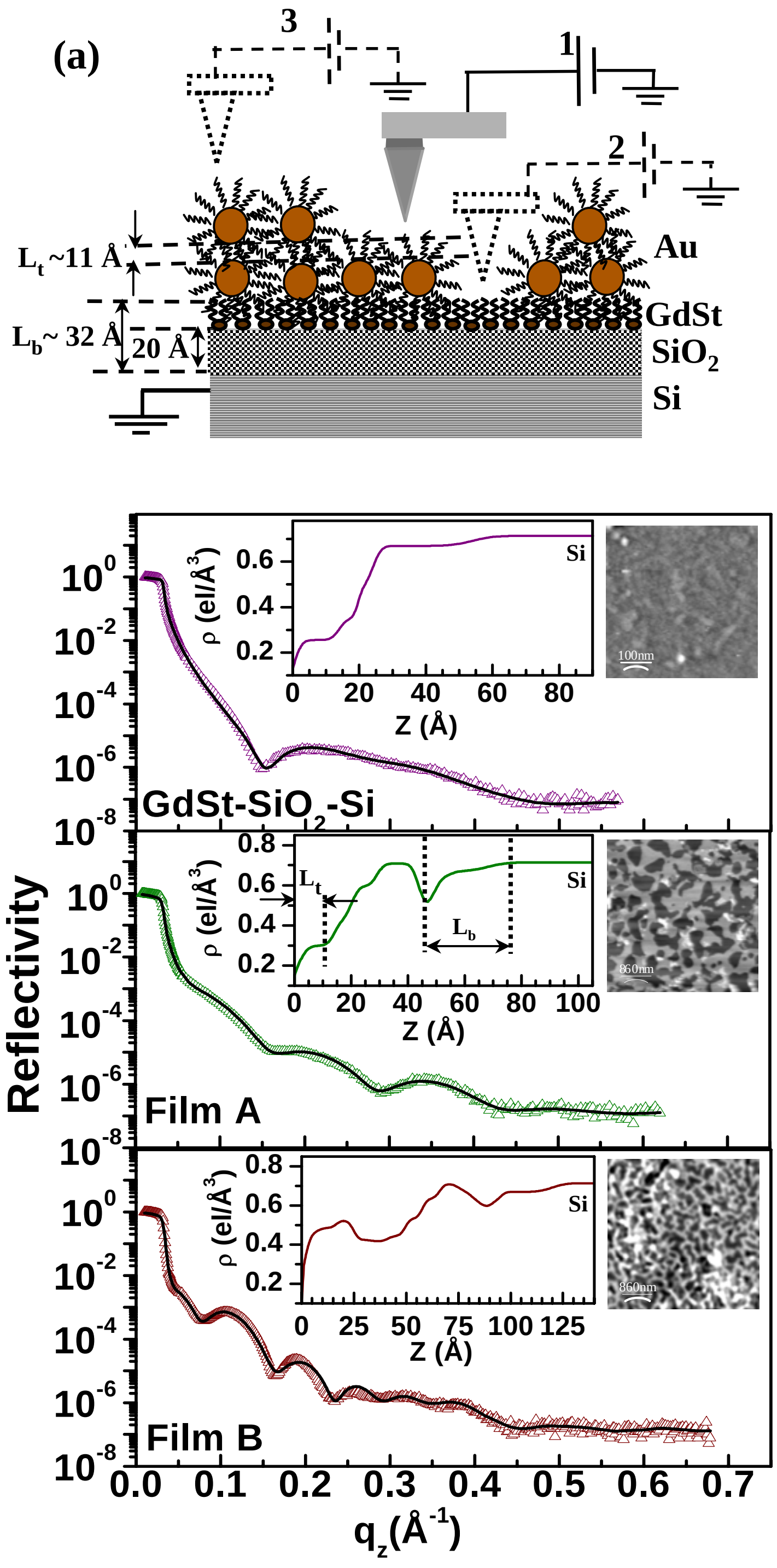

(b)

Fig. 1

Corresponding author: M. K. Sanyal

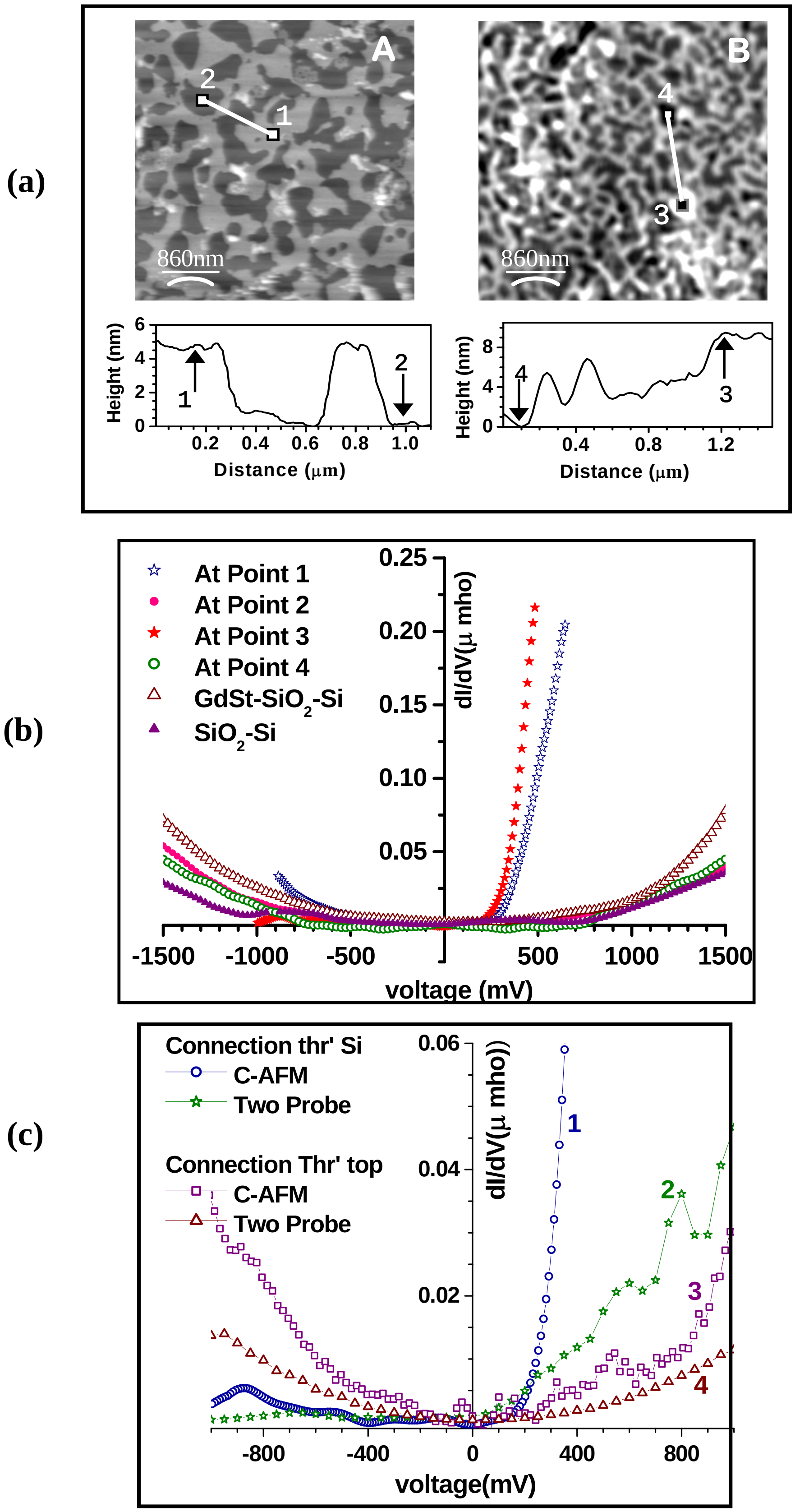

Fig. 2

Corresponding author: M. K. Sanyal

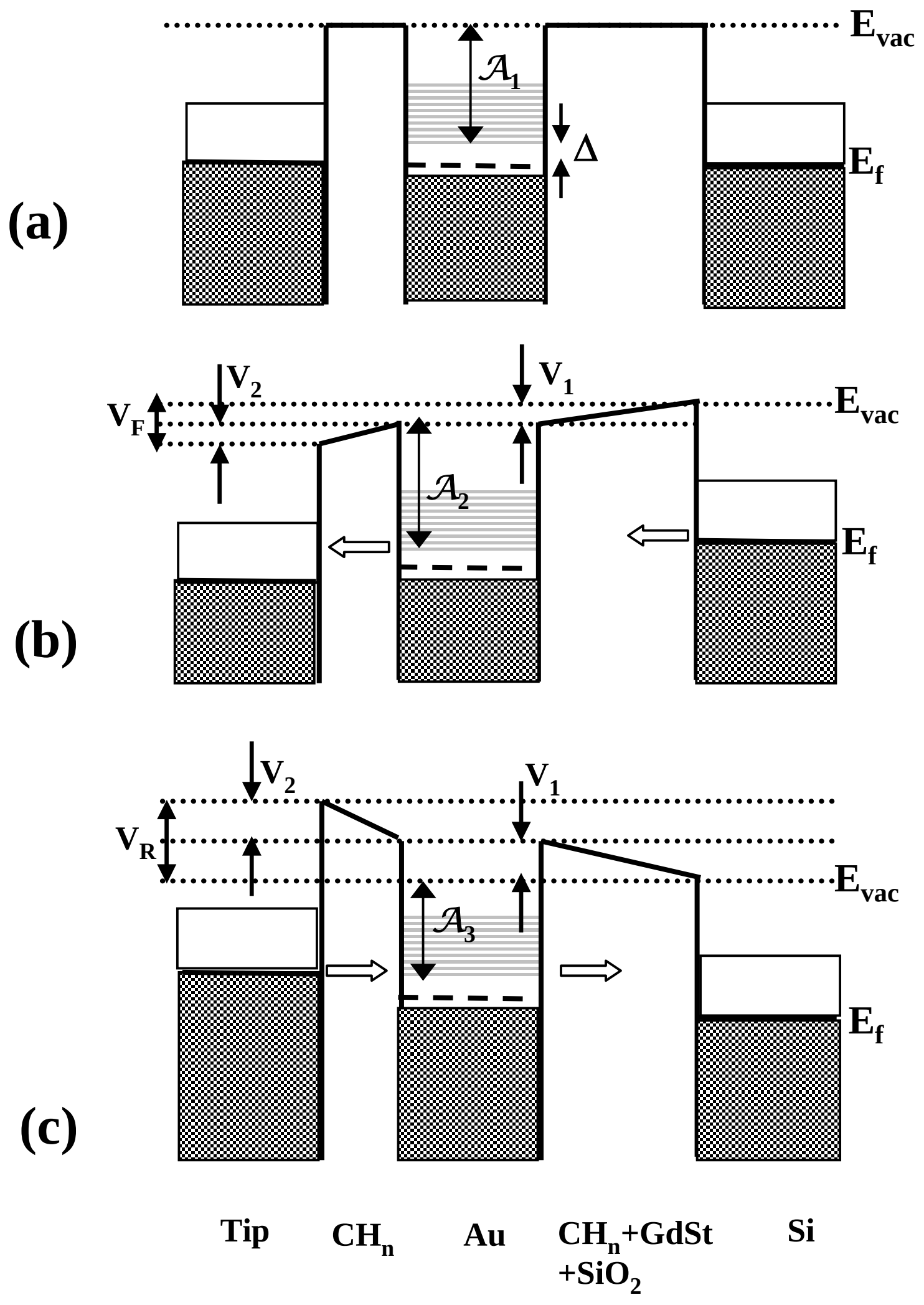

Fig. 3

Corresponding author: M. K. Sanyal